___
**TO QUOTE THIS WORK :**

V. Brien, B. Décamps. Low cycle fatigue of a nickel based superalloy at high temperature: deformation microstructures. Materials Science and Engineering: A, Elsevier, 2001, 316 (1-2), pp.18-31. ⟨doi.org/10.1016/S0921-5093(01)01235-7⟩. ⟨hal-02882485⟩

**THANK YOU**
___

# Low cycle fatigue of a nickel based superalloy at high temperature: deformation microstructures

by V. Brien [a] and B. Décamps [b]

[a][1]LSG2M, UMR 7584, Ecole des Mines de Nancy, Parc de Saurupt,
54042 Nancy Cedex, France
[b] Laboratoire de Chimie Métallurgique des Terres Rares, UPR CNRS 209,
2-8, rue Henri-Dunant, 94320 Thiais Cedex, France

**Abstract**

The microstructural characteristics of a single crystalline nickel based superalloy (AM1) tested under high temperature fatigue at 950°C are reported. For repeated fatigue ($R_\varepsilon=0$) through a range of cycle numbers N with imposed total strain amplitude $\Delta\varepsilon^t$ two main types of behaviour are found depending on N and $\Delta\varepsilon^t$. This allows a map of microstructures versus the number of cycles N and $\Delta\varepsilon^t$ to be constructed. A domain called A presents anisotropic microstructures due either to a partition of the plasticity throughout the γ channels, or to an oriented coarsening of the γ' precipitates of the so-called type *N* (rafts perpendicular to the loading axis). The domain called H shows homogeneous deformation microstructures. The coarsening observed in repeated fatigue is accounted for on the basis of a model proposed earlier in the literature by Véron. Alternate fatigue ($R_\varepsilon=-1$) also leads to the same type of coarsening, even for a very low number of cycles (N=113). The microstructures observed are ascribed to internal stress effects and are related to the possible shearing of the precipitates and to different diffusion processes.

[1] Corresponding author. V. Brien, Tel: +33-3 83 58 40 78; fax: +33-03 83 57 63 00.

E-mail address: brien@mines.u-nancy.fr

## 1. Introduction

Nowadays single crystal nickel based superalloys are commonly used in modern engines as turbine blades. As creep and thermo-mechanical behaviour in fatigue are the main types of loading of the alloy in service conditions, they have been extensively studied so far. However, the bottom of the blades are also exposed to low cycle mechanical repeated fatigue at high temperatures localised around 950°C. Although the macroscopic behaviour of the AM1 superalloy in fatigue has been extensively studied by [1-3] no careful study has been devoted to the elementary mechanisms of deformation. Actually, most of the available studies dealing with fatigue of single crystals of nickel-based superalloys are concerned with the alternate case ($R_\varepsilon$=-1). In these studies only, observations of a homogeneous deformation essentially located in the matrix has been reported [1,3,4-8].

In this present study, therefore, the objective was to investigate the microstructural behaviour of the AM1 superalloy during low cycle fatigue (LCF), during the setting up of microstructures. Repeated LCF ($R_\varepsilon$=0) matches more closely the real working conditions and was, as a consequence, first studied (cf. sections 3.1 and 4.1). The influence of the total imposed strain $\Delta\varepsilon^t$ ($\Delta\varepsilon^t = \varepsilon^{max} - \varepsilon^{min}$) is presented in section 3.1.4 while that of the number of cycles N on the microstructures is presented in sections 3.1.3 and 4.1.2. A map gathering all the $R_\varepsilon$=0 fatigue behaviours vs. the parameters $\Delta\varepsilon^t$ and N, within a realistic experimental range of values for both parameters, is presented at the end of section 3.1.4. Section 4.1 shows the importance of the role played by the mismatch at the γ/γ' interface in the building up of the deformation microstructure, and identifies the factors that govern the properties of the deformation map.

The effect of alternate fatigue ($R_\varepsilon$=-1) on the deformation microstructure is presented in sections 3.2 and 4.2. Simple compression is studied in section 3.2.1. The observed microstructure is analysed and compared to the one obtained in simple tension. This comparison is finally shown to be helpful for the understanding of the microstructural behaviour during alternate fatigue.

The materials, experimental procedures of the fatigue tests and the conditions for the transmission electron microscopy (TEM) are described in the first two sections.

## 2. Material and experimental procedure

*2.1. The superalloy AM1*

Due to their high temperature strength and stability, γ/γ' nickel-based superalloys are used as single crystal blades in gas turbine engines. The AM1 superalloy is a second-generation nickel-based single crystal alloy developed by SNECMA. Its nominal composition is Ni-5Co-3Cr-12.5Al-2.5W-2Ta-0.4Mo-1.2Ti (number indicate atomic %) [9]. The two phases γ and γ' constituting this alloy are coherent. The γ phase is a matrix of fcc structure and the γ' phase is made of ordered $L1_2$ (type $Cu_3Au$) cuboidal precipitates, representing about 70% of the total volume at room temperature. The γ' cuboids are aligned along cubic crystallographic directions as shown on the TEM micrograph of Fig.1. Their composition is very near that of the intermetallic phase $Ni_3Al$. The width of the matrix channels is about 100 nm and cuboids have a linear dimension of 400-500 nm. At 1223 K (950°C) the misfit of the alloy is negative and equal to $-10^{-3}$. It does not exhibit any tetragonal distortion [10]. The orientations of the tested specimens were determined by the X-ray back scattering Laue method and were found to deviate by less than 5° from the nominal [001]. Only one sample was found to deviate by 8°.

*2.2. Fatigue tests*

The type of fatigue tests and the experimental conditions were chosen in order to simulate the loading conditions of turbine blades knowing that these conditions are much more complex. As a consequence, isothermal total strain-imposed ($\Delta\varepsilon^t$) tests, $R_\varepsilon = \varepsilon^{min}/\varepsilon^{max} = 0$ (cf. 3.1), were performed at T=1223K (950°C), with a frequency f=0.25 Hz, the loading axis being taken parallel to the turbine blade main axis, [001].

All the tests were performed under air environment on INSTRON hydraulic tension machines. They were stopped at a number of cycles N or continued until failure. Each test is consequently defined by three parameters, i.e. N, $\Delta\varepsilon^t$ and the value of $R_\varepsilon$. In most of the tests, fatigue was repeated, i.e., $R_\varepsilon$=0. N has been varied from 1 to $2.10^4$ and $\Delta\varepsilon^t$ was in the range from 1.6 to 2.2%. The experimental details are given in Table 1. Two $R_\varepsilon$=-1 fatigue tests have been performed at $\Delta\varepsilon^t$=2 x 1.25 %= 2.50%. One of them was stopped after only one cycle, the other one was performed until rupture (N=113). The corresponding experimental details are given in Table 2. A single simple compression test has been performed at the same stress level as the simple tension test (738 MPa), with $\Delta\varepsilon^t$=1.73% and $\Delta\varepsilon^{pl}$=0.05% (the index pl refers to plastic deformation).

*2.3. TEM observations*

In the work presented here conventional TEM has been systematically used to study the deformation microstructures of the AM1 specimens after testing.

TEM observations were generally performed on {111} thin foils so that the observation plane is a slip plane which allows fine studies of the dislocation structure. This foil orientation is also convenient as the projections of the three cubic directions occur on it with a ternary symmetry. Some of the TEM observations have been done on {001} planes in order to examine interface dislocations and clearly visualise the dislocation activity within the channels. The thin foils were prepared by thinning down 100 μm thick discs by the double-jet polishing technique at room temperature, using the following electrolyte: 90% acetic acid and 10% perchloric acid. TEM examinations were performed on JEOL 2000EX, 200CX and PHILIPS CM20 electron microscopes. Micrographs were taken under bright field or weak beam conditions [11]. The FS/RH convention has been used [12] and the Burgers vectors were determined using the classical extinction criterion |**g**.**b**| = 0 [13].

The determination of the location of the extra plane associated with the edge component of a dislocation, in either the γ phase or the γ' phase, was systematically performed according to a rule deduced from [14] and [15]. If **g** is the diffraction vector and **b** the Burgers vector, such that |**g**.**b**|=2, the sense of **b** is given by the following rule: '**g**.**b**>0 if the black contrast is on the left side of the projection of the dislocation line' (the conventional orientation of the latter is from the bottom to the top of the micrograph). For the opposite **g**, the contrast will be shifted to the right, indicating that **g**.**b**<0. The strength and the asymmetry of the dislocation contrast obtained with |**g**.**b**|=2 allow the determination of the sense of b to be performed in a quite straightforward manner. Finally, the reproducibility of all the deformation microstructures, or configurations shown in what follows has been carefully checked.

## §3. Experimental Results
*3.1. Repeated fatigue, $R_\varepsilon$=0*
*3.1.1. Features common to all microstructures*

Strain is generally localised in the γ matrix phase. The nature of the active slip systems observed in all deformation microstructures with a [001] stress axis is consistent with the Schmid and Boas criterion i.e. with a preferential activation of 8 systems among the 12 {111} <110> possible ones. Other octahedral systems were found, however, but with a much smaller activity. Their occurrence can be explained by either local stress changes due to the accumulated deformation or by deviation of the experimental loading axis from the nominal axis.

Before describing the influence of N and $\Delta\varepsilon^t$, we present the microstructure of a test from the middle of the experimental $\Delta\varepsilon^t$ parameter range, stopped at N=1. The sample has been deformed until $\varepsilon^{max}$ and then unloaded to zero: this is a simple tension test.

It should be noted that a few $R_\varepsilon$=0 tests exhibit microstructures showing deformation bands. These bands have been shown to be features that are not specific of fatigue at this particular temperature.

*3.1.2. Microstructure of a simple tension test*

The alloy being monogranular and the loading axis being [001], the γ channels or corridors can have only two distinct orientations, parallel or perpendicular to the loading axis. For simplicity, the term vertical (respectively horizontal) refers in what follows to channels parallel (respectively perpendicular) to the loading axis.

The TEM. observations performed after the tension test ($R_\varepsilon$=0, $\Delta\varepsilon^t$=1.25%, N=1) show that only one type of γ channel is strained. Most of the dislocations are located in the {001} γ/γ' interfaces (cf. the configuration marked 1 in Fig. 2) i.e., only in interfaces perpendicular to the loading axis (whose projection is henceforth noted P on all micrographs). The dislocations form 4- or 6-sided polygonal networks. Their Burgers vector is always such that the extra-half plane associated to their edge component is located inside the precipitate phase (cf. Fig. 3). This is an evidence for the relaxation of the negative misfit existing between the two phases (it is recalled that the parameter of the γ phase is larger than that of the precipitate). The analysis of the mobile dislocations in the corridors show that they are of screw character. When moving, they leave edge or 60° mixed dislocations at the γ/γ' interfaces. As these interfacial dislocations come from the activation of different slip systems, their intersection and reaction produce new dislocation segments, thereby reducing the total elastic energy. The polygonal networks thus created, commonly described as honeycombs in the literature, have been currently observed at the interfaces of different nickel based superalloys after creep [17-19].

In contrast, the vertical interfaces are totally dislocation-free and undeformed (cf. mark 3 in Fig. 2). This microscopic plastic anisotropy of the different types of channels under an applied stress has already been observed by other authors in the MC2 alloy by *in situ* experiments at 1223 K [20] and in the AM1 alloy [19].

*3.1.3. Influence of the number of cycles N on the deformation microstructure*

A second cycle does not change the deformation microstructure much with respect to the previous description, except that the channels perpendicular to the loading axis exhibit a higher density of dislocations. When the number of cycles increases, the deformation spreads out significantly into the two vertical channels (cf. Fig. 4 where N=25). Although different slip systems seem to be activated, the density of dislocations is nevertheless not high enough to allow for the occurrence of interfacial networks.

If N is further increased (N=1300) and still for $\Delta\varepsilon^t$=1.25%, the microstructural anisotropy is still present but it is much less marked. Networks appear in the {010} and {100} interfaces, so that most of the observed channel areas look rather uniformly deformed (cf. Fig. 5). The determination of the extra half-planes associated with the edge component of the {001} interfacial dislocations show that most of them are located in the γ' phase. A few exceptions were encountered for large cycle numbers (N>25). The study was not performed for {010} or the {100} interfaces.

In summary, the main effect of increasing N, given the total strain amplitude $\Delta\varepsilon^t$, is to reduce the anisotropy of the microstructure through an increased deformation of the channels containing the loading axis. Actually, it was found that the effect of N depends on the value of $\Delta\varepsilon^t$ (cf. next section).

*3.1.4. Influence of the imposed total strain $\Delta\varepsilon^t$ on the deformation microstructures*

A comparison of the deformation microstructures obtained with different values of $\Delta\varepsilon^t$ was performed at a fixed value of N = 200 cycles. When $\Delta\varepsilon^t$ is small (0.6%), the deformation microstructure is fully anisotropic, with a very well marked initiation of type N coarsening (see Fig. 6). At $\Delta\varepsilon^t$ =1.50%, the extension of the plastic deformation towards the vertical channels is very significant and the anisotropy has disappeared. No coarsening is observed and the microstructure is similar to the one of Fig. 5. For a larger value of $\Delta\varepsilon^t$ (2.2%) the deformation microstructure is again completely homogeneous. No difference can be noticed among the different types of interfaces and some shearing of the precipitates is evidenced by the presence of superdislocations separated by antiphase boundaries [21]. Studies like the one presented in Fig. 3 show that dislocations indifferently relax or not the misfit in the vertical interfaces for high $\Delta\varepsilon^t$.

Therefore, the higher the total imposed strain amplitude , the smaller the number of cycles for which the anisotropy disappears. For very small values of $\Delta\varepsilon^t$, this vanishing of the anisotropy is not observed at all. On the contrary, the microstructure evolves from a micro-mechanical anisotropy to a morphological anisotropy, as the γ' precipitates coarsen into rafts perpendicular to the [001] loading axis ($R_\varepsilon=0$, $\Delta\varepsilon^t=0.6-1.3\%$, $N\approx10^4$) showing this way that diffusion effects are strong under such conditions.

A few sheared configurations are also observed for high values of the total imposed deformation.

The map shown in Fig. 7 summarises the type of microstructure observed as a function of the two parameters N and $\Delta\varepsilon^t$. In this boundary, two domains appear that are labelled A (for anisotropic) and H (for homogeneous).

### 3.2. Alternate fatigue $R_\varepsilon=-1$
#### 3.2.1. Microstructure after a simple compression test

In this case, the dislocations are essentially located in the {010} and {100} channels and at the corresponding interfaces. Furthermore only one slip system is found to be active in each type of channel (cf. Fig. 8). As a consequence, no polygonal networks like those observed in $R_\varepsilon=0$ fatigue can be seen. They are replaced by regular sets of parallel dislocation segments left at the interfaces by the dislocations of the active slip system. Such experimental observation has also been made by [22]. No shearing of the γ' phase has been observed in this type of test.

#### 3.2.2. Microstructure after an alternate test (the very beginning of fatigue, N=1)

The microstructure obtained after alternate loading (cf. Fig. 9-a) can be considered as stemming from the addition of the microstructure due to simple tension (Fig. 2) and of the microstructure due to simple compression (Fig. 8). Some shearing of the γ' phase also occurs, according to a mechanism that seem to produce extrinsic stacking faults (only a few have been studied here, cf. Fig. 9-b). This type of fault is generally expected for this alloy after dynamic compression [23]. Studies performed on interfacial dislocations in the vertical channels show that these dislocations are actually relaxing the interfacial misfit.

#### 3.2.3. Microstructure after an alternate test, at rupture (N=113)

Thin foils extracted from the specimen at some distance from the fracture surfaces were prepared with orientations perpendicular to either [001] or to [111]. The TEM observations show that the precipitates have drastically coarsened in rafts perpendicular to the loading axis [001] (cf. Fig. 10-a). Thus, the $R_\varepsilon=-1$ fatigue produces marked type N coarsening after a very small number of cycles.

## §4. Discussion
### 4.1. Repeated fatigue, $R_\varepsilon=0$
#### 4.1.1. Internal stresses and deformation microstructures

Although preferential localisation of the plastic stain was expectable in the very ductile γ matrix, geometrical differentiation was not trivial to predict. Usually, in monocrystalline alloys deformed in fatigue, clustering of dislocations originates from multiple interactions between dislocations, according to the "forest" mechanisms. Here, the presence of the precipitates strongly modifies this behaviour and localises the plastic deformation in the channels. On top of this, comes a geometrical differentiation of plasticity reflecting an heterogeneity of the local stresses due to the presence of internal stresses. These are geometry-dependent. They are the misfit stresses coming from the lattice mismatch between the two phases (cf. §2.1) and the compatibility stresses coming from the strong similarity of the superalloy and a composite.

#### 4.1.1.1. The misfit and the deformation microstructures

The misfit stresses (cf. section 2.1) have been the object of many studies and of some calculations by the Finite Element Method. Royer, Bastie, Bellet and Strudel from [10] have shown that the misfit of the AM1 alloy is negative at 1223 K (950°C). The stresses $\sigma_\delta$ resulting in the γ phase of an unloaded superalloy are then compressive in all channels along the two perpendicular directions parallel to the interfaces. When the sample is loaded in tension with an applied stress, $\sigma_a$, the local stress is smaller than $\sigma_a$ in the vertical channels, since $\sigma_\delta$ directly reduces $\sigma_a$. This is not so in the horizontal channels. As a result, the latter are easier to deform than the vertical ones. As it was shown in the present work that the dislocations located at the γ/γ' interfaces relax the misfit, a simple estimate is performed below to show how many of them are needed to completely relax the misfit. The corresponding number of interfacial dislocations segments can be calculated by making use of an expression due to Brooks [24]. This equation gives the number of pure edge dislocations necessary to completely relax the misfit in two perpendicular directions:

$$\sqrt{2}\ d_{\gamma'}\cdot\delta\ /\ b = (\sqrt{2}\ (400\ \text{to}\ 500)\ \times 10^{-3}\ )/\ ((\sqrt{2}/2)\ \times 3.58 \times 10^{-1}) = 2\ \text{to}\ 3\ \text{dislocations}$$

$d_{\gamma'}$ is the linear dimension of the γ' precipitates (400 to 500 nm), δ is the absolute value of the misfit between the two phases ($10^{-3}$), b is the modulus of the Burgers vector of the a/2 [110] dislocation and a is the lattice parameter of the γ phase.

Thus, only 4 to 6 dislocations are necessary to totally relax the misfit on the {001} interfaces. The corresponding critical number of cycles, $N_O$, should consequently be very small. Although it is very difficult to estimate experimentally when the misfit is effectively reduced, $N_O$ can be estimated to be in the range between 1 and 25 (cf. figs. 3 and 5), depending on the value of $\Delta\varepsilon^t$. Indeed, the plastic deformation per cycle is higher for larger $\Delta\varepsilon^t$, so that the smaller is $\Delta\varepsilon^t$, the larger $N_O$ must be. When the number of cycles increases beyond $N_O$, a reverse of the misfit stresses can be expected since most of the interface dislocations still have their extra half-plane in the γ' phase. Some cases of non relaxation of the misfit have indeed been recorded for large N (N≥1300). It seems also, according to the present experimental results, that the extension of the deformation to the vertical {100} and {010} channels starts when the critical number of cycles $N_O$

*4.1.1.2. The compatibility stresses and the deformation microstructure*

In a composite material where two phases are strained in parallel, internal stresses appear as a result of the requirement of strain continuity at the interface between the phases. These stresses reduce the applied stress in the softer phase and increase it in the harder one. The peculiar geometrical configuration of the superalloy, which is made up of γ' cuboids aligned along <001> and loaded along [001] can be considered as a composite made up of vertical "fibres". So, as soon as the deformation of the vertical channels is allowed (upper right part of Fig. 7), compatibility stresses will appear. Since they reduce the applied stress in the softer phase, they reduce the slip activity in the vertical channels. As a consequence, they slow down the transition to the domain H.is reached. This suggests that resolved shear stresses on activated slip systems in the two kinds of channels have the same values when N> $N_O$.

*4.1.2 Influence of the number of cycles N on the deformation microstructures: diffusion effects, by-passing, coarsening*

After a larger number of cycles ($\Delta\varepsilon^t$=1.5%, from N=200, test duration=13 min 20s), the influence of diffusion processes is clearly evidenced. Lower densities of dislocations are observed, indicating that dislocation recovery is effective. In addition, directional coarsening of precipitates occurs, for large values of N and small values of $\Delta\varepsilon^t$. Such effects explain why the interfacial networks achieve large sizes, despite the high cumulated plastic deformation (cf. table 2). It is of course not surprising to observe that diffusion effects are of great importance here, since the test temperature of 1223K corresponds to $0.7T_m$, where $T_m$ is the melting temperature of the AM1 alloy. However, the present observations prove that diffusion processes occur very early in $R_\varepsilon$=0 fatigue. The directional coarsening noticed here is of the same type as the one typically observed in tension crept superalloys. As a matter of fact, precipitates coarsen in rafts perpendicular to the loading axis. This is what is called type N coarsening [25]. Type N coarsening describes a coarsening of the precipitates in plates perpendicular to the loading axis. The occurrence of this type of coarsening can be explained using the model proposed by Véron [19] for nickel based superalloys. This model is based on the fact that the diffusion flux is directly related to the gradients of elastic energy between the interfaces. This elastic energy is modified by a relaxation of the γ/γ' misfit. A relaxed interface will grow at the expense of the non relaxed ones. Given the value of the misfit of the AM1 superalloy, an interface will be relaxed if the extra-half plane associated with the edge component of the interfacial dislocation is located in the γ' phase. As the experimental investigations show, the horizontal interfaces are relaxed for small N. Over N=25, it is not systematic even if the interfaces seems to be relaxed in majority. No investigation has been performed on the vertical ones. But even though all the dislocations locating at the vertical interfaces were relaxing, their density is not large enough in the vertical interfaces in comparison to the one in the horizontal channels to eliminate the elastic gradients. This implies that horizontal interfaces are earlier and more relaxed than the vertical ones. The gradients are then such that they provoke a type N coarsening.

The directional coarsening, perpendicular to the cubic loading axis, observed in $R_\varepsilon$=0 fatigue at small $\Delta\varepsilon^t$ is then nothing else than a consequence of the anisotropy and the nature of the deformation at the precipitates scale.

*4.1.3. Influence of the total imposed strain $\Delta\varepsilon^t$ on the deformation microstructures*

*4.1.3.1. Influence of $\Delta\varepsilon^t$ on shearing*

In the present work, no example was found of precipitate shearing creating super stacking faults like those observed by [23] in a γ' phase. However, for large values of the total strain ($\Delta\varepsilon^t$ =2.2%) the shearing by pairs of 1/2[110] dislocations have been noted to be significant. In such conditions, the applied stress was 975MPa. It follows that, at the testing temperature, a transition is observed in the mechanism by which precipitates are passed by the dislocations, from by-passing to shearing by superdislocations. This shearing mechanism directly induces a reduction of the anisotropic character of the deformation microstructure.

*4.1.3.2. Influence of $\Delta\varepsilon^t$ on diffusion processes, and coarsening*

The higher is the applied stress, the greater are the climb forces on dislocations [26-28]. Dislocations and interfaces provide diffusion short-cuts and, as a matter of fact, help dislocations to by-pass the precipitates by climb. On the other hand, recovery by climb accounts for the lowest dislocations densities recorded with larger values of $\Delta\varepsilon^t$, for a given number of cycles N. If one considers a dislocation relaxing an horizontal interface climbing around a precipitate from the horizontal channel to the vertical one: one can quickly see considering the geometry, that this dislocation is also relaxing the vertical interface. It implies that climb is a factor of reduction of the elastic gradients. It follows that the main effect of diffusion, for large values of $\Delta\varepsilon^t$, is to destroy the anisotropic character of the deformation microstructure and provokes the vanishing of the elastic gradients between the vertical and the horizontal interfaces. No physical reason then exists any longer for producing oriented coarsening according to Véron's model [18].

**§ 5. Expectations on fatigue behaviour at other temperatures**

The experimental data have been obtained for 1223K (T= 950°C). If the temperature of the test is modified the relative importance of the heterogeneity of plastic flow in the matrix structure, the shearing of the precipitates and the diffusion processes will be modified.

At low temperatures, it is expected that precipitate shearing occurs in a significant manner, as found in the AM1 superalloy in tension at the same temperature [17] and as reported in the literature [1,4]. At very low temperatures (T<<0.6Tm), no coarsening should be observed.

At high temperatures, one can expect diffusion processes to be stronger. Climb of dislocations should move the domain H towards the left of the map of Fig. 7, due to an easier slip spreading towards the vertical channels. The domain of coarsening should also be shifted in the same manner.

At intermediate temperatures, if precipitate shearing is not possible, the domain of coarsening and the domain H are expected to be shifted towards the right of the map for $R_\varepsilon$=0 fatigue.

**§ 6. General conclusions**

The following remarks and conclusions can be drawn from the present study.

• As a result of systematic TEM investigations a map of deformation microstructures in repeated fatigue ($R_\varepsilon$=0) has been constructed. This map presents two main domains corresponding to 1) a mechanically induced anisotropy of the microstructure (domain A) and to 2) an homogeneous microstructure (domain H).

- The influence of the two parameters N, the number of cycles, and $\Delta\varepsilon^t$, the total imposed deformation which governs the stress level, have been pointed out.
- The work stresses also the important role played by internal stresses on the building up of the initial deformation microstructure and on its evolution. The different plasticity mechanisms explaining the micro-mechanical superalloy response have been identified. The mechanisms leading to the evolution of the microstructures from the domain A to the domain H are (i) - the reset of $\sigma_\delta$ in the horizontal channels, (ii) - diffusion-assisted climb and precipitate by-passing, and (iii) – the shearing of the precipitates at high stresses. Low strain amplitudes are present in the A domain. Inside this one, type N coarsening appears at high number of cycles. Véron's model (1995) can then be used to explain the type N rafted structure observed in $R_\varepsilon=0$ fatigue at such a temperature.

For large values of $\Delta\varepsilon^t$, no anisotropy is observed and three-directionnal coarsening occurs. Our view is that either the stress is so high that precipitate shearing currently occurs, dislocations coming from the horizontal channels being then transferred into the vertical ones, or the stress is so high that climb forces on dislocations contribute substantially to the spreading of the deformation towards the vertical channels. It could also be both.
- Under the light of this work on fatigue at 1223 K (950°C), some expectations at other temperatures are listed for the low cycle fatigue behaviour of nickel-based superalloys.
- A parallel has been drawn between the microstructures observed in compression and in tension, indeed, it is almost opposite in form between compression and tension.
- Alternate fatigue is shown to induce a three-dimensional distribution of the plastic deformation in the γ channels, since tension activates plastic flow in the channels perpendicular to [001] and compression in the others. Alternate fatigue provokes type N coarsening of the superalloy at very small numbers of cycles. However, the present study does not bring any further information regarding either the nature of the coarsening or the high speed at which it occurs. Further experimental investigations would be necessary to validate Véron'model in alternate fatigue.

This work underlines the strong importance of the initial spatial distribution of plastic flow due to the spatial distribution of internal stresses. In [29] a detailed study of this behaviour is performed, based on the evolution of the internal stresses.

**Acknowledgements**
This work has been supported by the "CPR Stabilités Structurales des superalliages monocristallins" Snecma and the CNRS. The as cast materials were provided by Snecma, which is acknowledged for its interest in this work and for having performed all the rupture tests in $R_\varepsilon=0$ fatigue. The authors wish to thank Dr. Luc Rémy (Ecole des Mines-Evry-France) for helpful discussions and for providing fatigue test facilities. They wish to thank more especially Dr. A.J.Morton (CSIRO-Melbourne-Australia) for subsidising, providing us with fatigue facilities and for having welcomed us at the CSIRO. The work was performed at LMS, URA CNRS 1107, Université Paris-Sud, 91405 Orsay Cedex and at LEM, Unité mixte CNRS-ONERA UMR 0104, BP 72, 92322 Châtillon Cedex, in France.

**FIGURE CAPTIONS**

Figure 1: Transmission electron micrograph showing the microstructure (non deformed) of nickel based superalloy AM1 after an homogenization treatment ensuring the uniform distribution of the precipitates and the absence of dislocations at interfaces.

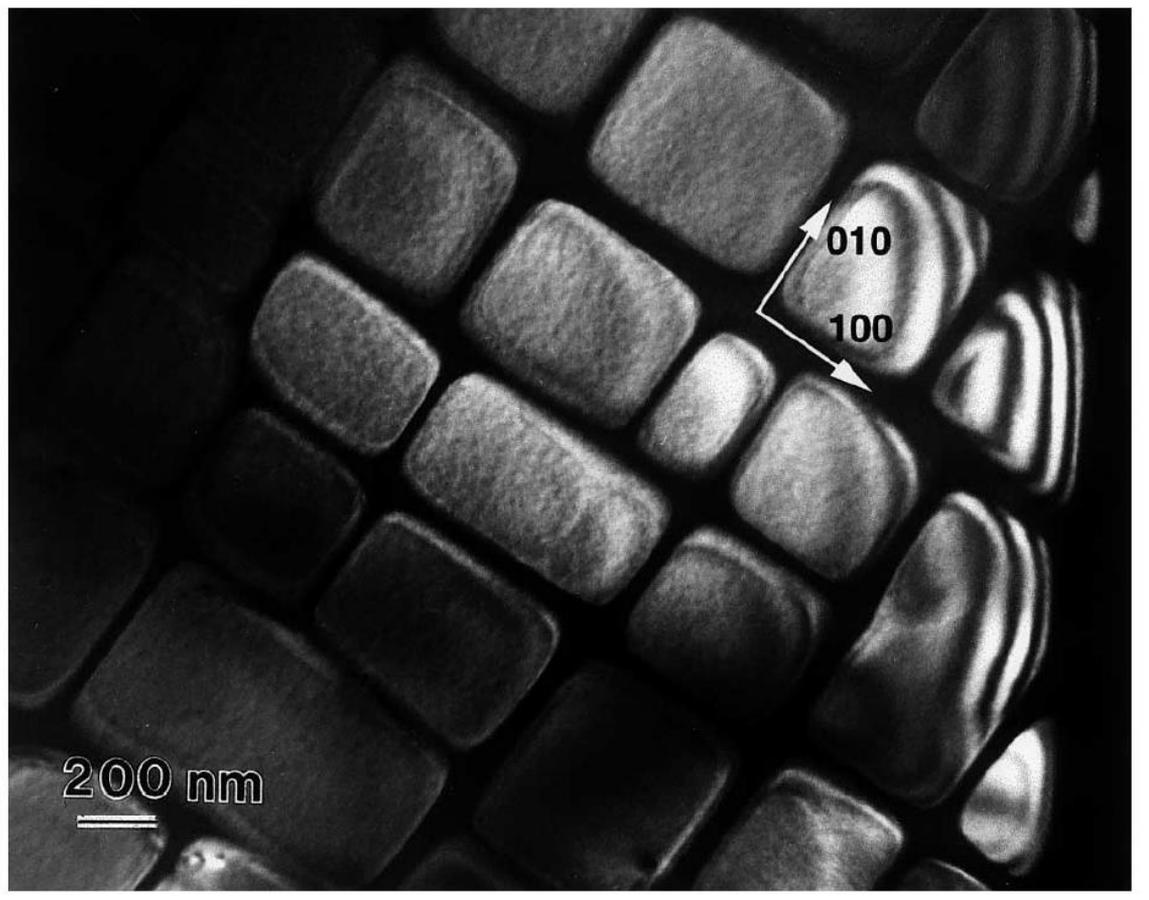

Figure 2: ($R_\varepsilon=0$, $\Delta\varepsilon^t=1.27\%$, N=1), Zone axis = [111] allowing an equal projection of the three channels. P is the projection of the loading axis. Dislocations are localized at {001} interfaces, perpendicular to the loading axis: cf. arrows 1. {010} and {100} channels are empty of dislocations: cf. arrows 2.

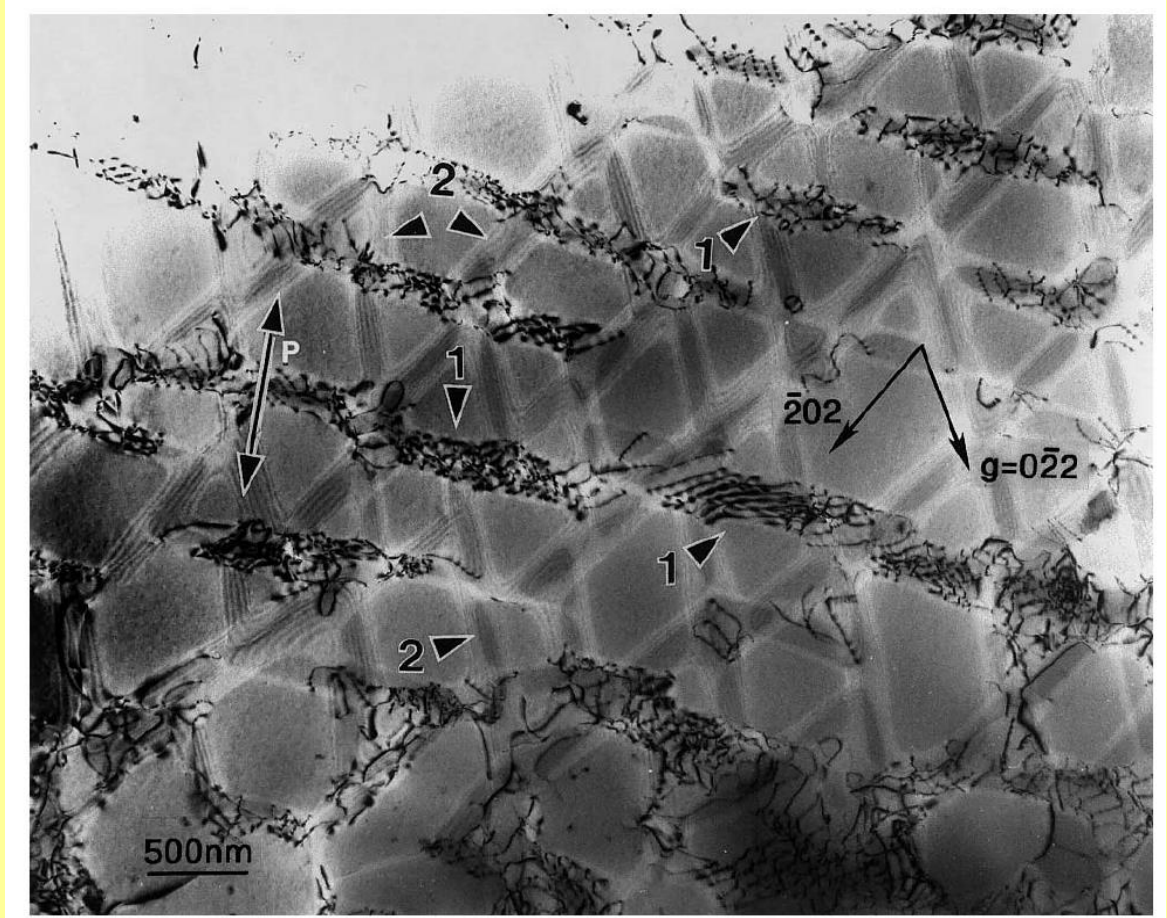

Figure 3: ($R_\varepsilon=0$, $\Delta\varepsilon^t =1.27\%$, N=1) Network localized at a {001} interface: (a) Network entirely in contrast: weak beam image; zone axis = [111], (b) Dislocations A fill the condition |g.b|=2. Their extra plane associated to their edge component is localized in $\gamma$' ; zone axis = [111], (c) Schema of the network. Extra planes are under the diagram and the left view of the cut of the foil at the interface.

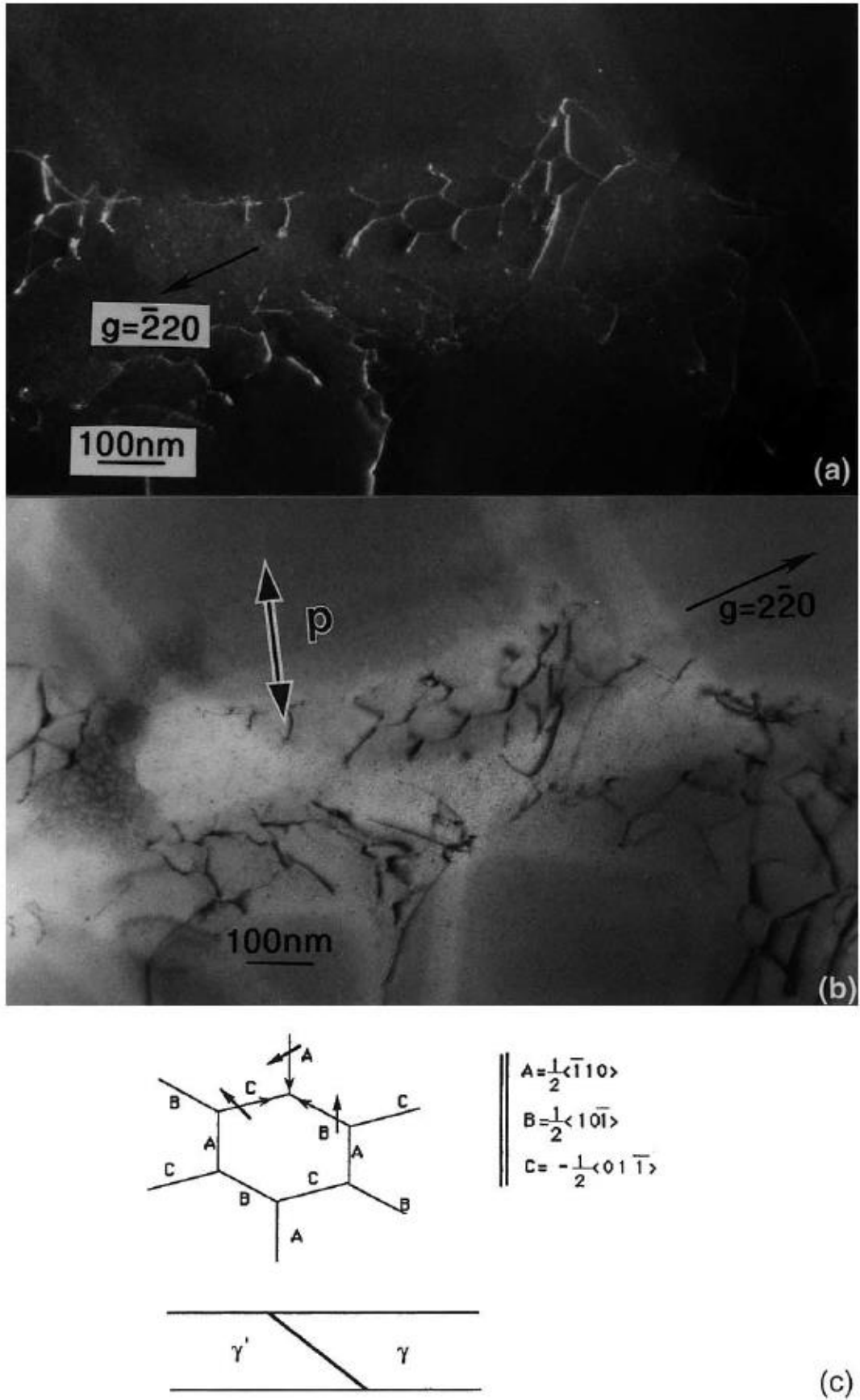

Figure 4: ($R_\varepsilon=0$, $\Delta\varepsilon^t =1.25\%$, N=25) Zone axis = [111]. The channel perpendicular to the loading axis whose projection is still p is the most deformed one. Other channels contain at this level a noteworthy amount of deformation. 1 shows dislocations alone in the last channels.

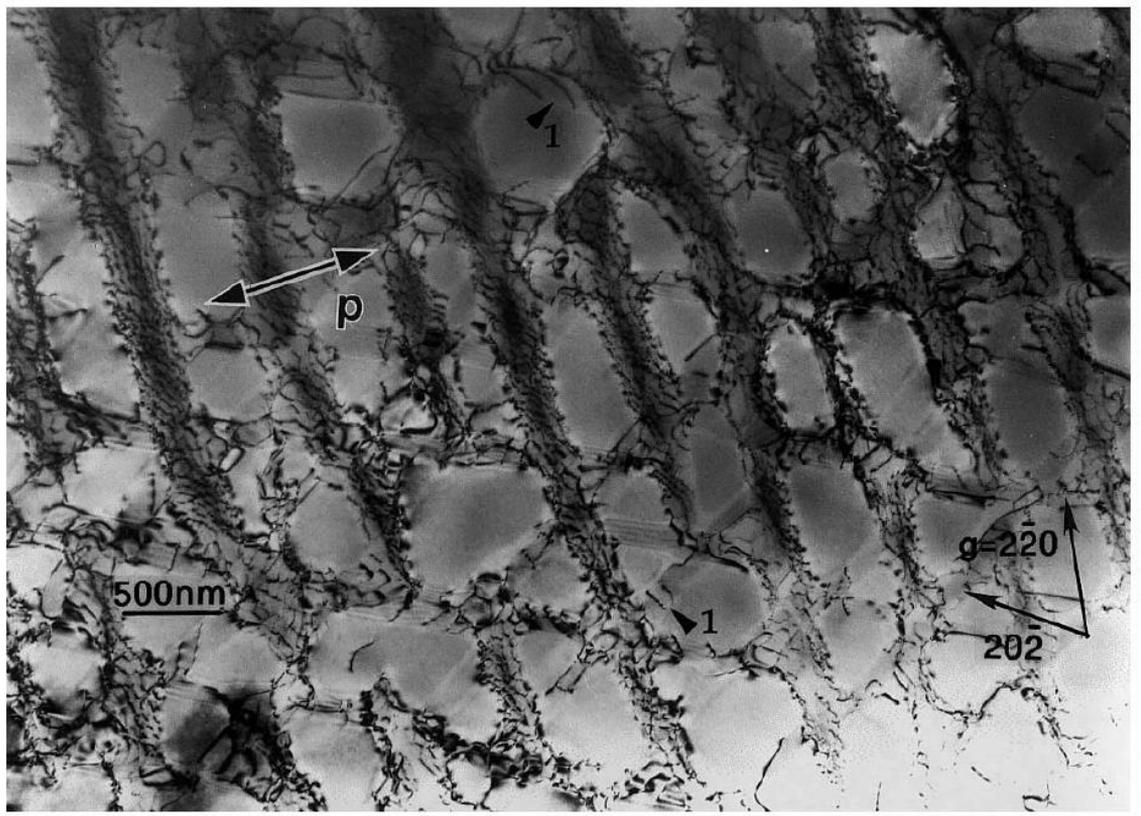

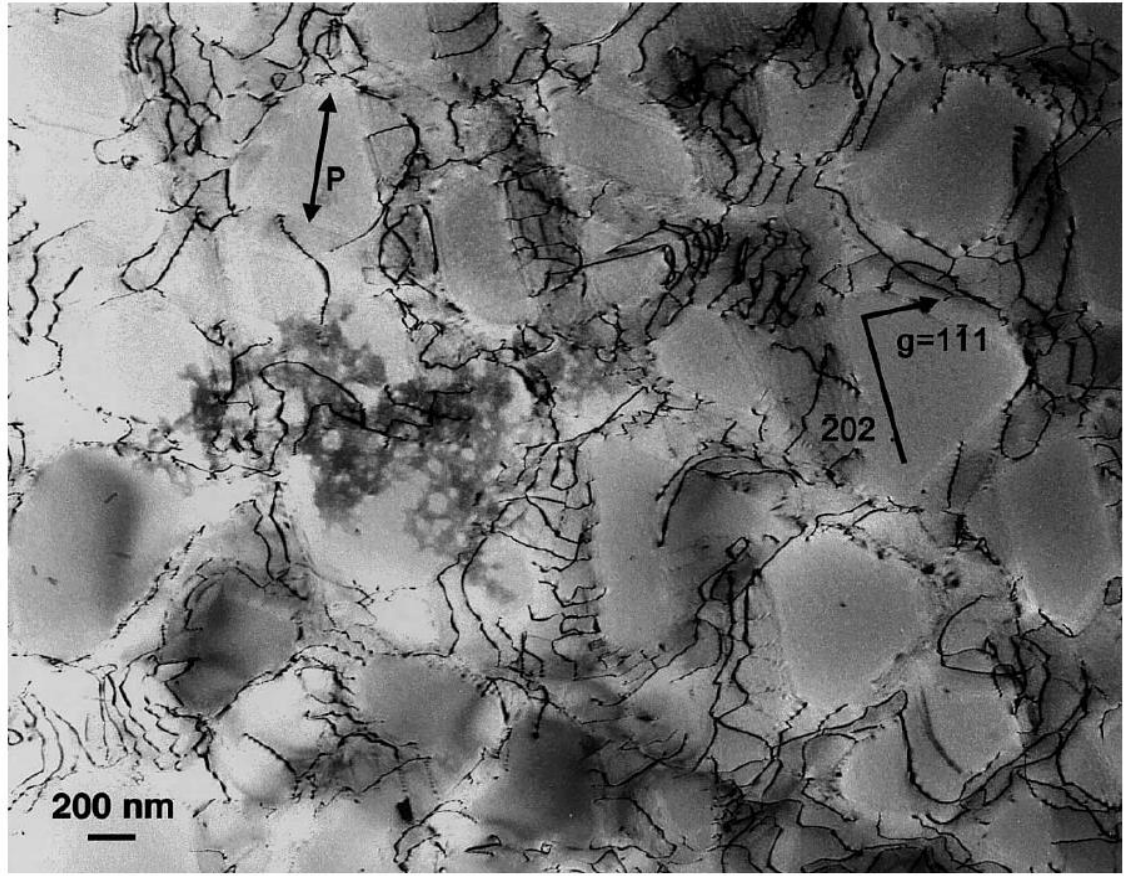

Figure 5: ($R_\varepsilon=0$, $\Delta\varepsilon^t =1.28\%$, N=1300). Zone axis = [121]. P is the projection of the loading axis. Homogeneous deformation microstructure.

Figure 6: ($R_\varepsilon=0$, $\Delta\varepsilon^t =0.63\%$, N=200). (a) Zone axis = [112]. Concentration of the deformation only in channels perpendicular to the loading axis ; bright field image, (b) Zone axis = [110]. Type N coarsening: perpendicular to [001]; dark field micrograph with a substructure spot.

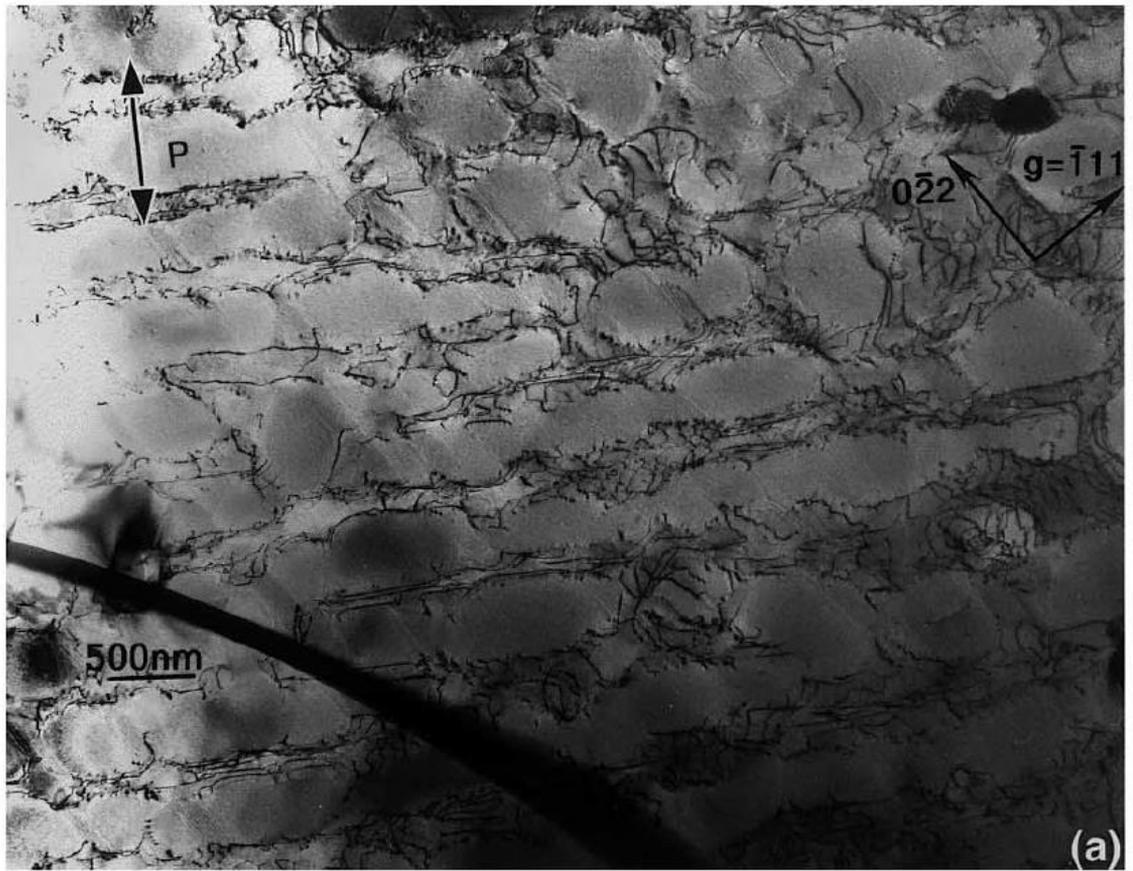

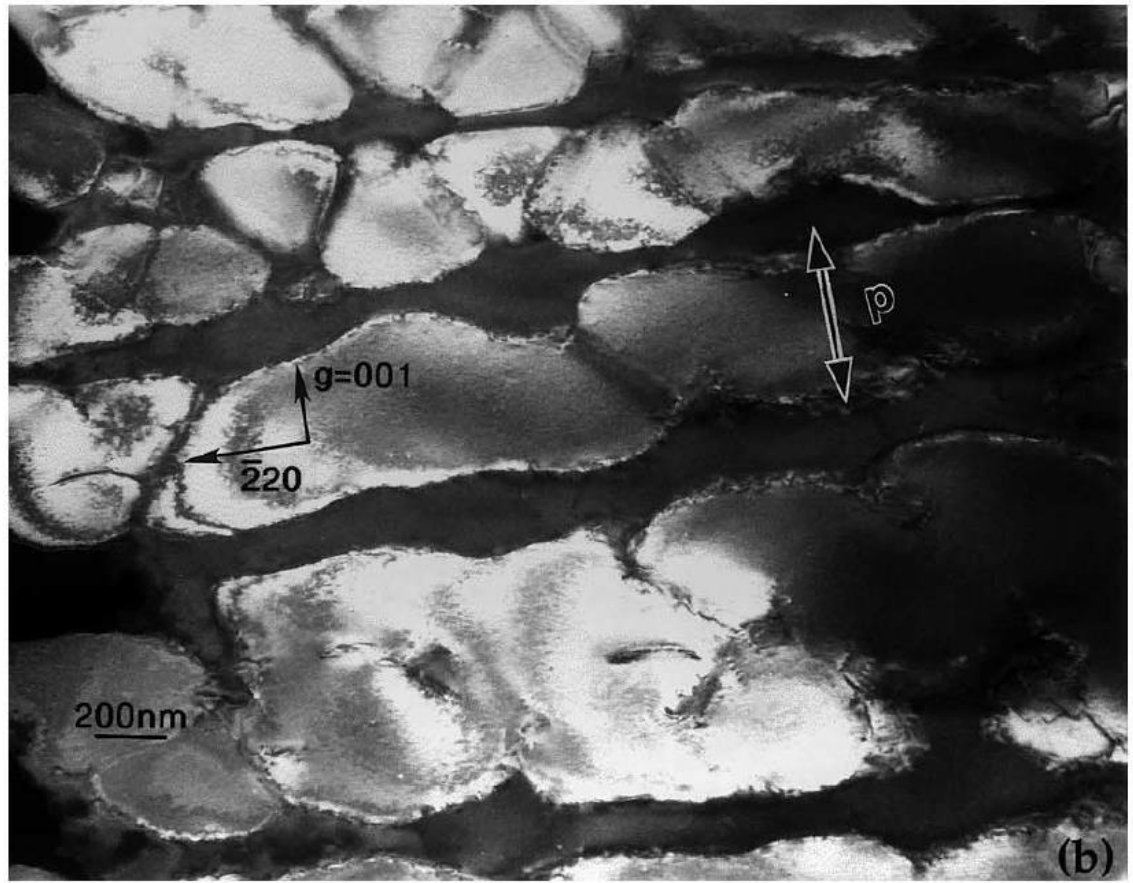

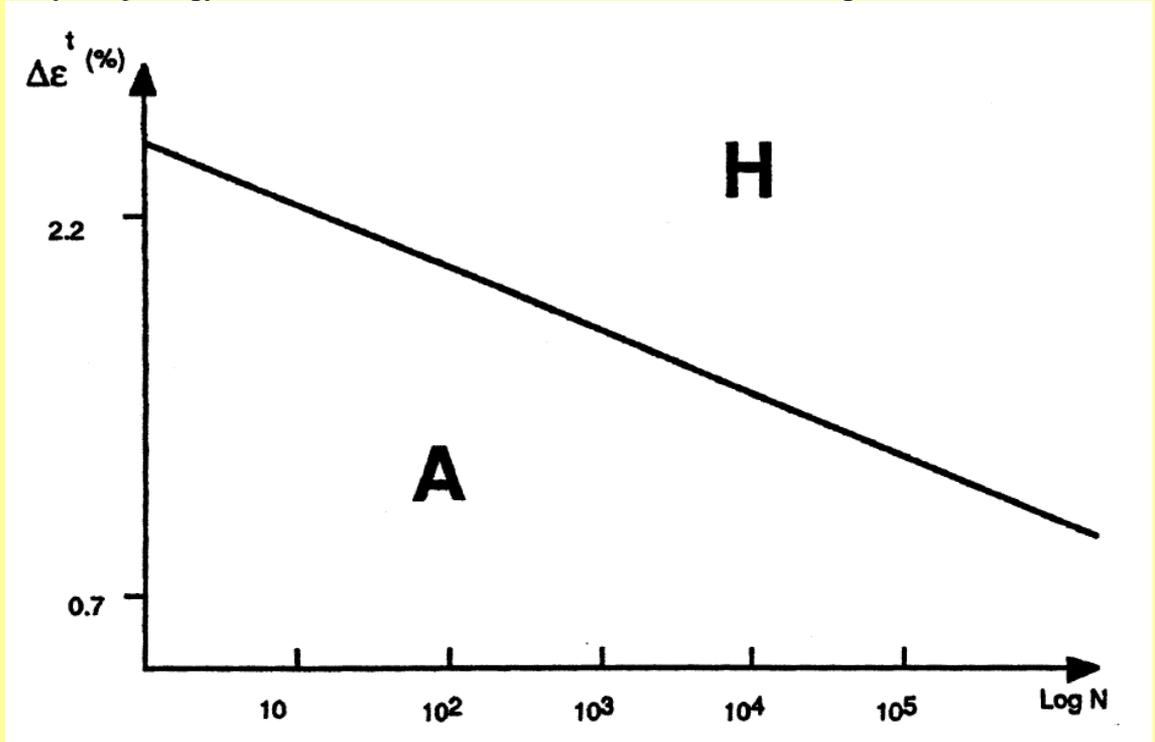

Figure 7: Map of $R_\varepsilon =0$ deformation microstructures. Appearance of two main domains: A and H. In A, the microstructures of deformation are anisotropic (either micromechanically, or by morphology), in H, the microstructures of deformation are homogeneous.

Figure 8: ($R_\varepsilon=0$, $\Delta\varepsilon^t =1.73\%$, N=1, compression test): zone axis = [111]. Deformation is localized only in channels parallel to the loading axis [001]. Every type of channels contains only one type of slip system.

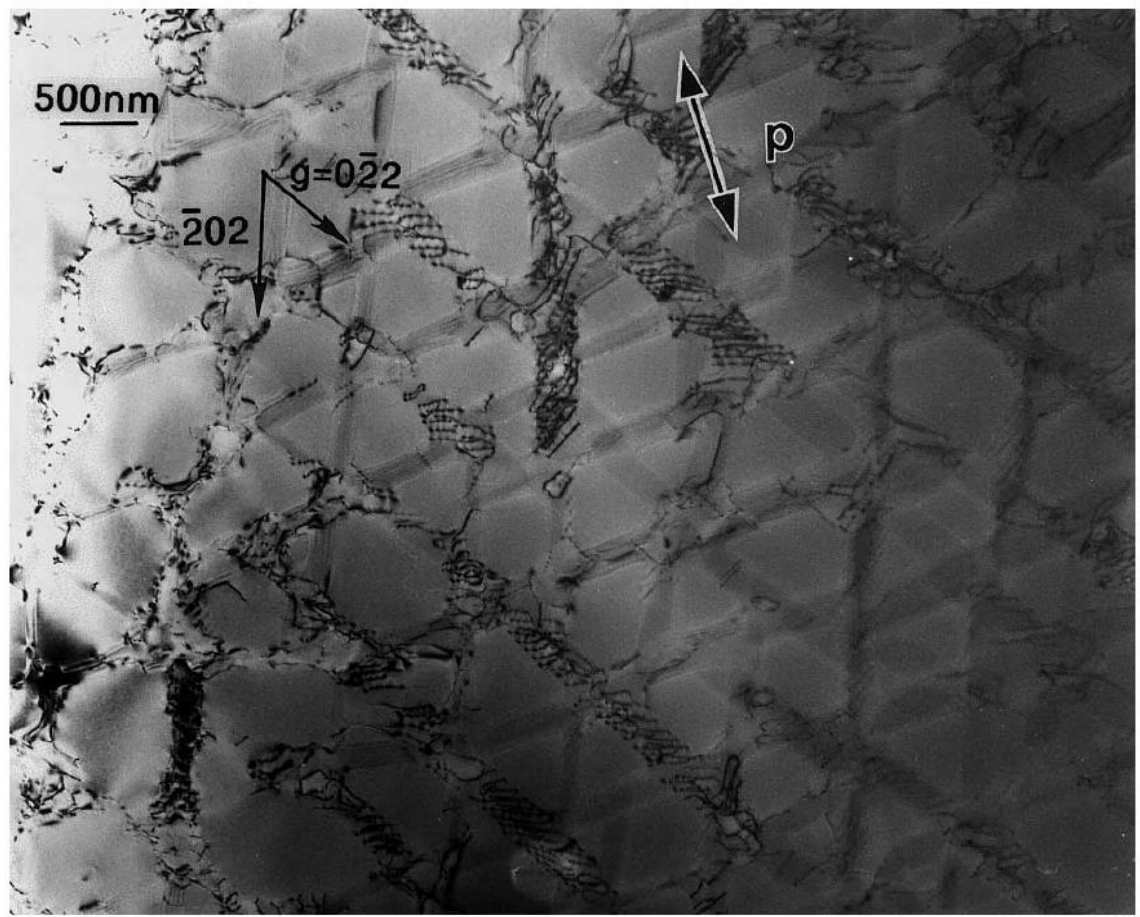

Figure 9: ($R_\varepsilon= -1$, $\Delta\varepsilon^t =1.26\%$, N=1). (a) horizontal channels are concerned by deformation as well as in tension. The new fact coming from the compression is that vertical channels contain one slip system per type of channel as in simple compression: zone axis = [112]. (b) Shearing of precipitates is noted by a mechanism presented by Décamps et al. 1991. This one is of extrinsic nature located in {111} plane: zone axis = [112].

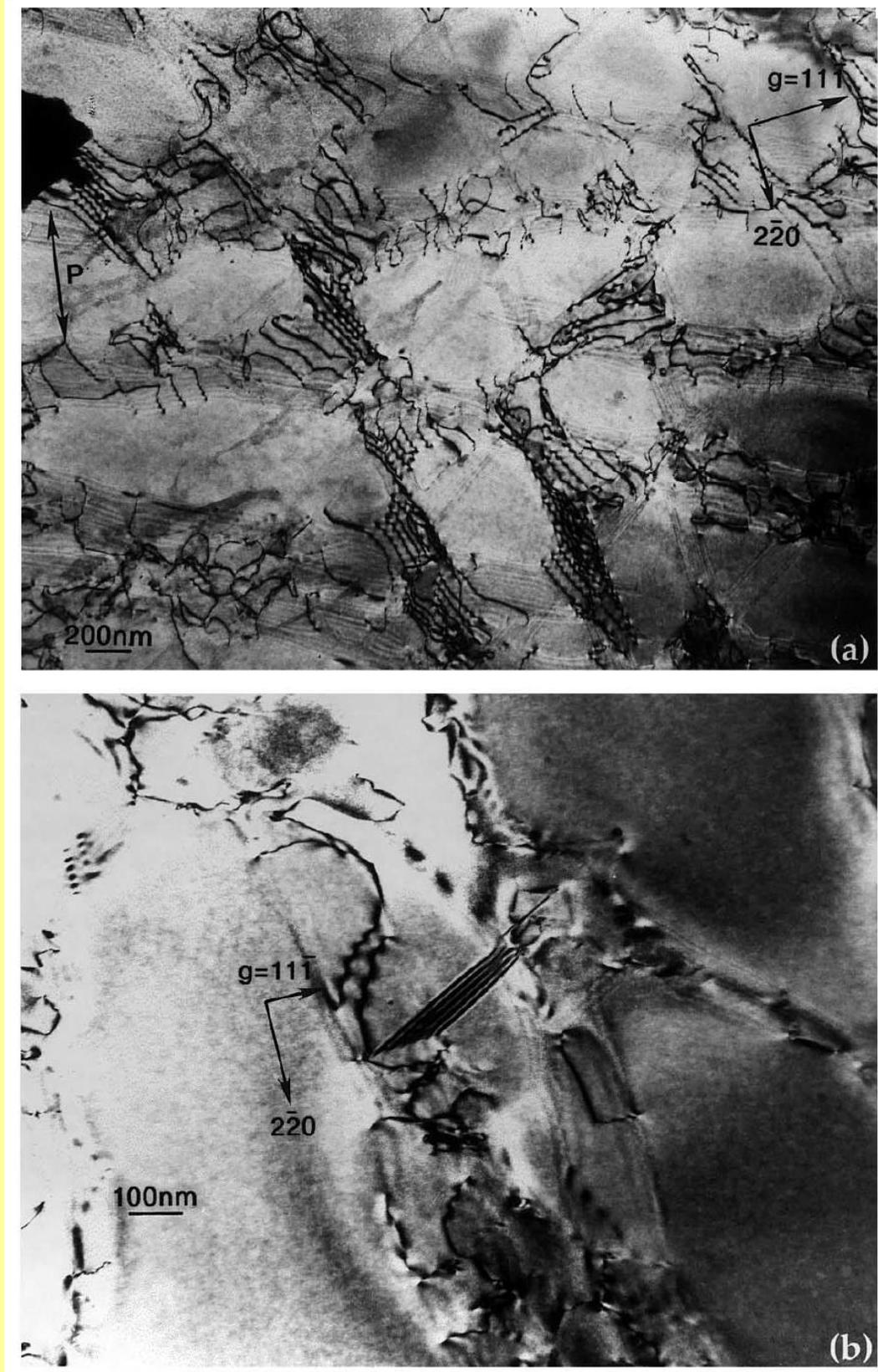

Figure 10: ($R_\varepsilon$= -1, $\Delta\varepsilon^t$ =1.26%, N=113, broken test). (a) Strong and very extended coarsening of type N is noted: zone axis = [001] ; thin foil perpendicular to [001] ; (b) loop of dislocation has been noted often at intersection of channels in the γ phase ; thin foil perpendicular to [001] ; zone axis = [112].

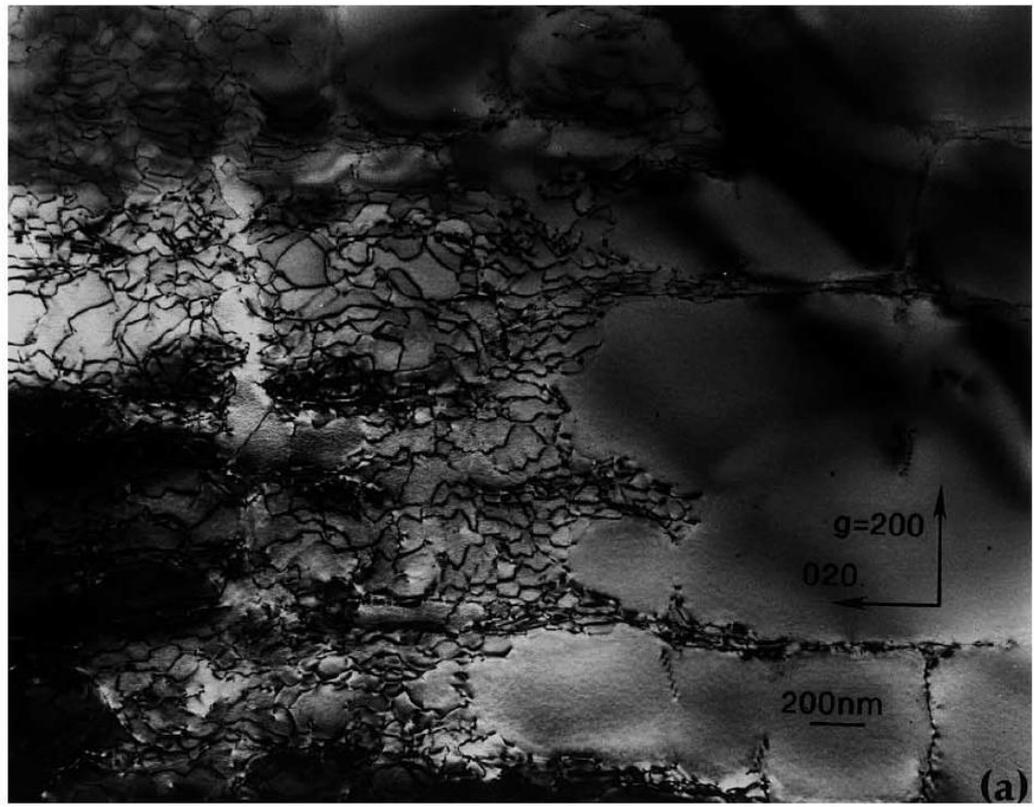

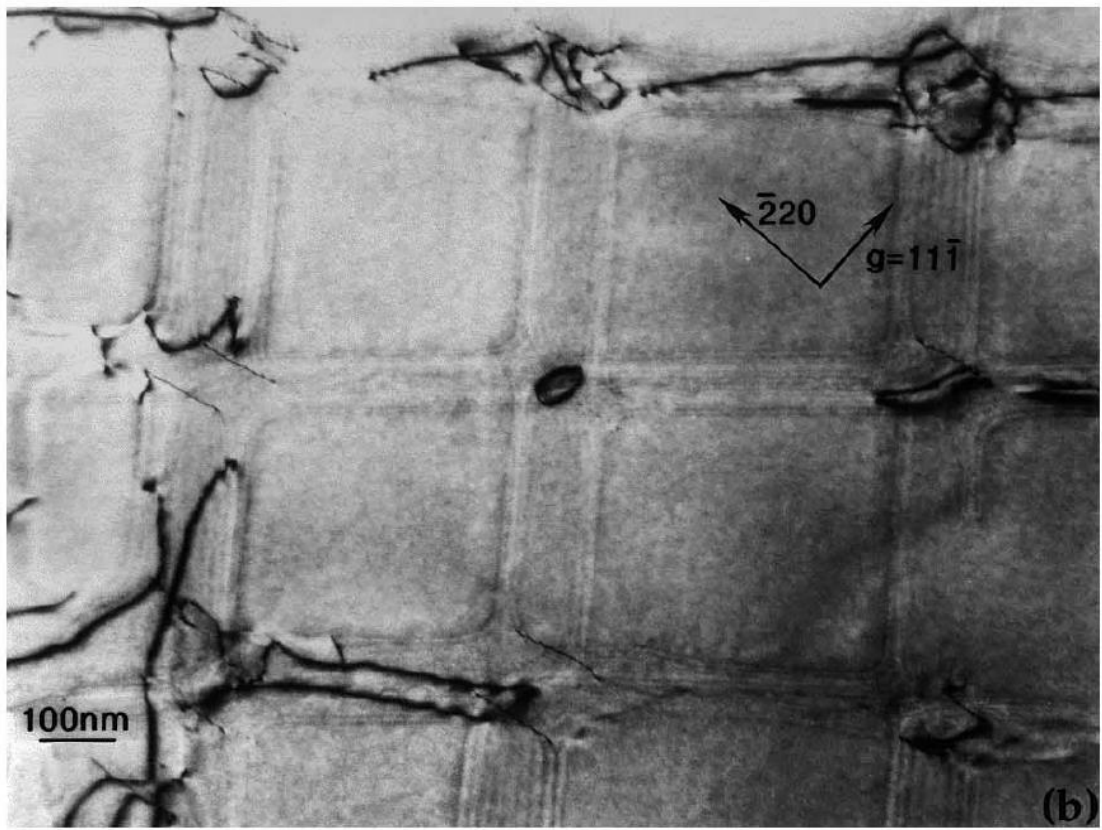

**TABLES**
Table 1: Mechanical data of $R_\varepsilon = 0$ tests.

| Δε tot. % | Cycle / Total number of cycles | Maximal Stress in tension $\sigma_{tens}^{max}$ MPa | Maximal Stress in compression $\sigma_{comp}^{max}$ MPa | Average plastic deformation per cycle % | Total plastic deformation % |
|---|---|---|---|---|---|
| 0.63 | 1<br>200 | 458<br>415 | 9<br>43 | 0.01 | 2 |
| 0.70 | 1<br>1/2 lifetime<br>5630 | 608<br>428 | -<br>178 | 0.04 | 200 |
| 0.79 | 1<br>1/2 lifetime<br>17117 | 663<br>438 | -<br>212 | 0.06 | 1027 |
| 0.79 | 1<br>1/2 lifetime<br>21314 | 642<br>395 | -<br>277 | 0.06 | 1279 |
| 1.09 | 1<br>1/2 lifetime<br>3345 | 815<br>447 | -<br>453 | 0.12 | 401 |
| 1.27 | 1<br>1 | 739 | - | 0.153 | 0.153 |
| 1.33 | 1<br>25 | 661<br>637 | 143<br>157 | 0.247 | 0.47 |
| 1.25 | 1<br>2 | 732<br>632 | 57<br>151 | 0.16 | 2.10 |
| 1.26 | 1<br>1300 | 759<br>497 | 47<br>238 | - | 125 |
| 1.28 | 1<br>1300 | 785<br>451 | 50<br>357 | - | 125 |
| 1.6 | 1<br>5<br>200 | 704<br>663<br>520 | 39<br>76<br>205 | 0.80 | 160 |
| 1.6 | 1<br>5<br>25 | 707<br>670<br>620 | 56<br>81<br>126 | 0.80 | 20 |
| 2.2 | 1<br>5<br>200 | 975<br>910<br>610 | 85<br>150<br>470 | 1.10 | 100 |

Table 2: Mechanical data of $R_\varepsilon = -1$ tests.

| $\Delta\varepsilon^{tot}/2$ % | Cycle<br>Total number of cycles | $\sigma_{tens}^{max}$<br>MPa | $\sigma_{comp}^{max}$<br>MPa | Plastic deformation per cycle<br>% | Plastic cumulated deformation<br>% |
|---|---|---|---|---|---|
| 1,26 | 1 | 932 | 952 | (+) 0.253<br>(-) 0.449<br>(+) 0.111 | 0.81 |
| 1,26 | 1<br>Stabilized<br>113 | 960<br>874 | 887<br>887 | 0,77 | 85 |